\documentclass[preprint, amsmath, amssymb, preprintnumbers, superscriptaddress, floatfix, showpacs, showkeys, aps, prl]{revtex4-1}

\usepackage{graphicx}
\usepackage{float}
\usepackage{color}

\begin{document}

\title{Z$_2$ topology and superconductivity from symmetry lowering of a 3D Dirac Metal Au$_2$Pb}

\author{Leslie M. Schoop}
\email{lschoop@princeton.edu}
\affiliation{Department of Chemistry, Princeton University, Princeton New Jersey 08544, USA.}
\author{Lilia S. Xie}
\affiliation{Department of Chemistry, Princeton University, Princeton New Jersey 08544, USA.}
\author{Ru Chen}
\affiliation{Molecular Foundry, Lawrence Berkeley National Laboratory, Berkeley, California 94720, USA.}
\affiliation{Department of Physics, University of California, Berkeley, CA 94720, USA.}
\author{Quinn D. Gibson}
\affiliation{Department of Chemistry, Princeton University, Princeton New Jersey 08544, USA.}
\author{Saul H. Lapidus}
\affiliation{X-ray Science Division, Advanced Photon Source, Argonne National Laboratory, Argonne, IL 60439, USA.}
\author{Itamar Kimchi}
\affiliation{Department of Physics, University of California, Berkeley, CA 94720, USA.}
\author{Max Hirschberger}
\affiliation{Joseph Henry Laboratory, Department of Physics, Princeton University, Princeton New Jersey 08544, USA.}
\author{Neel Haldolaarachchige}
\affiliation{Department of Chemistry, Princeton University, Princeton New Jersey 08544, USA.}
\author{Mazhar N. Ali}
\affiliation{Department of Chemistry, Princeton University, Princeton New Jersey 08544, USA.}
\author{Carina A. Belvin}
\affiliation{Wellesley College, Wellesley, MA 02481, USA.}
\affiliation{Research Experience for Undergraduates Program, Princeton Center for Complex Materials (PCCM), Princeton, NJ, 08544, USA.}
\author{Tian Liang}
\affiliation{Joseph Henry Laboratory, Department of Physics, Princeton University, Princeton New Jersey 08544, USA.}
\author{Jeffrey B. Neaton}
\affiliation{Molecular Foundry, Lawrence Berkeley National Laboratory, Berkeley, California 94720, USA.}
\affiliation{Department of Physics, University of California, Berkeley, CA 94720, USA.}
\affiliation{Kavli Energy Nanosciences Institute, Berkeley, California 94720, USA.}
\author{N. P. Ong}
\affiliation{Joseph Henry Laboratory, Department of Physics, Princeton University, Princeton New Jersey 08544, USA.}
\author{Ashvin Vishwanath}
\affiliation{Department of Physics, University of California, Berkeley, CA 94720, USA.}
\author{R. J. Cava}
\affiliation{Department of Chemistry, Princeton University, Princeton New Jersey 08544, USA.}

\date{\today}

\maketitle


\textbf{3D Dirac semi-metals (DSMs) are materials that have massless Dirac electrons and exhibit exotic physical properties \cite{wang2012dirac,liu2014discovery,wang2013three,PhysRevLett.113.027603,neupane2014observation,liang2014ultrahigh}. It has been suggested that structurally distorting a DSM can create a Topological Insulator (TI) \cite{wang2012dirac,wang2013three}, but this has not yet been experimentally verified. Furthermore, quasiparticle excitations known as Majorana Fermions have been theoretically proposed to exist in materials that exhibit superconductivity and topological surface states \cite{fu2008superconducting}. Here we show that the cubic Laves phase Au$_2$Pb has a bulk Dirac cone above 100 K that gaps out upon cooling at a structural phase transition to create a topologically non trivial phase that superconducts below 1.2 K. The nontrivial Z$_2$ = -1 invariant in the low temperature phase indicates that Au$_2$Pb in its superconducting state must have topological surface states. These characteristics make Au$_2$Pb a unique platform for studying the transition between bulk Dirac electrons and topological surface states as well as studying the interaction of superconductivity with topological surface states.}

Since their prediction and subsequent discovery, DSMs have emerged as new type of electronic material with relativistic, massless electrons in the bulk; they can indeed be thought of as 3D analogues of graphene \cite{wang2012dirac,liu2014discovery,wang2013three,PhysRevLett.113.027603,neupane2014observation}. Two recent materials, Cd$_3$As$_2$ and Na$_3$Bi, have been shown by photoemission to be 3D DSMs with Dirac points protected by crystalline symmetry, and further transport experiments have shown properties such as ultrahigh mobility and large linear magnetoresistance \cite{liu2014discovery,PhysRevLett.113.027603,neupane2014observation,liang2014ultrahigh}. The effects of structural and electronic instabilities on the 3D Dirac spectrum are not well known to date, however. Breaking crystalline symmetry, for example, can gap out the Dirac spectrum, adding a finite mass term and creating a topological insulator \cite{wang2012dirac,wang2013three}. Furthermore, the behavior of Dirac electrons in 3D materials under the influence of electronic or magnetic instabilities such as ferromagnetism and superconductivity is not yet well explored. 

Cubic Laves phases have the general formula X$_2$Y, where the X atoms form a pyrochlore lattice and the Y atoms form a diamond lattice. Several cubic laves phases are superconducting and temperature dependent structural phase transitions of cubic laves phase are also common \cite{roy1992ceru2}.
Materials with pyrochlore lattices are of particular interest because they can display exotic properties such as frustrated magnetism, and are the proposed basis for topological electronic states \cite{snyder2001spin,wan2011topological}. To date, however, cubic Laves phases based on the pyrochlore lattice have not been investigated as potential 3D DSMs.

Here we report the properties and electronic structure of Au$_2$Pb, which is a cubic Laves phase \cite{Au2Pbstructure} that has been reported to superconduct below 1.2 K \cite{hamilton1965some}. Our electronic structure calculations indicate that cubic Au$_2$Pb has a bulk Dirac cone  at room temperature. However, likely due to the presence of other electron and hole pockets, we find through high resolution X-ray diffraction experiments that on cooling Au$_2$Pb undergoes a previously unreported symmetry breaking phase transition near 50 K that gaps out the Dirac spectrum in the high temperature phase. The result is a low temperature nontrivial massive 3D Dirac phase with Z$_2$ = -1 topology. Upon further cooling, Au$_2$Pb becomes superconducting. This implies that Au$_2$Pb is a bulk superconductor with spin polarized topological surface states whose existence is guaranteed by the nontrivial Z$_2$ topology of the low temperature phase. This is of particular interest as proposals for realizing Majorana fermions include inducing the effect of superconductivity on spin polarized surface states. Our results show that the surfaces of single crystals of Au$_2$Pb are a natural platform for realizing Majorana fermions and for probing other consequences of the interaction of topological surface states with superconductivity.
 

Several mm sized single crystals of Au$_2$Pb were grown out of Pb flux. Single crystal diffraction analysis at T = 100 K confirmed the previously reported cubic Laves phase crystal structure of Au$_2$Pb (figure 1 (a)). The superconducting transition was also confirmed with resistivity and specific heat measurements. The resistivity shows a sharp drop at T$_c$ = 1.3 K. The superconducting transition was further examined through temperature-dependent measurements of the electrical resistivity under applied magnetic field. In figure 2 (a) the T$_c$ is plotted versus the applied field. A clear linear dependence of $\mu_0$H$_{c2}$(T) is seen; a linear fit to the data reveals  dH$_{c2}$/dT$_c$ = - 0.0928 T/K. By using the Werthamer-Helfand-Hohenberg (WHH) relationship \cite{werthamer1966temperature} $\mu_0$H$_{c2}$(0) = -0.7T$_c$ dH$_{c2}$/dT$_c$, we estimate the zero-temperature upper critical field to be 83 mT. With this information, the coherence length can be calculated by using the Ginzburg-Landau formula $\xi_{GL}$(0) = ($\phi_0$/2$\pi$H$_{c2}$(0))$^{1/2}$, where $\phi_0$=h/2e and is found to be $\xi_{GL}$(0) = 629 \AA. Both values are comparable to the ones measured in elemental Pb.

Figure 3 presents the electronic part of the specific heat of Au$_2$Pb. A sharp anomaly is observed at the superconducting transition temperature (1.19 K) indicating bulk superconductivity. The low temperature heat capacity above T$_c$ follows the expected C$_V$ = $\gamma$T + $\beta$T$^3$ relation, where $\gamma$T describes the electronic contribution to the heat capacity and $\beta$T$^3$ the phonon contribution, which can be related to the Debye temperature $\Theta_D$. $\gamma$ is the Sommerfeld parameter. By fitting the specific heat from 1.3 - 2.5 K we obtain $\gamma$ = 2.2 mJ/mol$\cdot$ and $\Theta_D$ = 139.6 K. By estimating the magnitude of the specific heat jump with the equal area approximation we obtain $\Delta$C/$\gamma$T$_c$ = 1.95. This is significantly higher than the BCS value of 1.43, but lower than the value measured in elemental Pb \cite{neighbor1967specific}. With these results and assuming $\mu^*$ = 0.13, the electron-phonon coupling constant ($\lambda_{ep}$) can be calculated from the inverted McMillan's formula \cite{mcmillan1968transition}: $\lambda_{ep} = \frac{1.04~+~\mu^*\ln \left(\frac{\Theta_D}{1.45T_c}\right)}{\left(1-0.62\right)\mu^*\ln \left(\frac{\Theta_D}{1.45T_c}\right)~-~1.04}$ and is found to be 0.58, which suggests weak coupling. Having the Sommerfeld parameter and the electron-phonon coupling, the non-interacting density of states at the Fermi energy can be calculated from: \textit{N}(E$_F$) = $\frac{3\gamma}{\pi^2k_B^2\left(1 + \lambda_{ep}\right)}$. The value obtained for Au$_2$Pb, \textit{N}(E$_F$) = 0.6 states eV$^{-1}$ per formula unit, agrees well with the $\approx$ 0.5 states eV$^{-1}$ per formula unit predicted by band structure calculation (see below). Table S2 summarizes the superconducting properties. The analysis of the superconducting properties indicates that Au$_2$Pb is a robust weakly coupled BCS superconductor. This is in contrast to Cu$_x$Bi$_2$Se$_3$ where the inhomogeneity, present due to the Cu dopant, affects the superconducting properties \cite{hor2010superconductivity}.

 The full temperature range of the resistivity is shown in in figure 2 (b). The data reveals several discontinuities, which we determined are related to structural phase transitions; one of them (at 50.4 K) is also seen in the heat capacity data (figure 3). In agreement with these observations, our low temperature synchrotron powder diffraction data shows that Au$_2$Pb undergoes two structural phase transitions on cooling. It is cubic above 100 K, below that temperature an unknown intermediate structure appears for a short range of temperature. Then on cooling to 55 K and below, a third structure appears. This structure we identify as being primitive orthorhombic; below 40 K this is the only phase remaining. We were able to fully solve the structure (R$_{Bragg}$ value of 5.73\%) in the space group \textit{Pbcn} with the lattice constants a = 7.9040(7) \AA ~b = 5.5786(10) \AA~ and c =11.1905(2) \AA. Table S1 summarizes the atom positions. This low temperature structure is a slightly distorted version of the cubic Laves phase structure, with the pyrochlore planes being buckled (see figure 1).

The electronic structures of the cubic and orthorhombic phase structures for Au$_2$Pb are shown in figure 4. Counting all the parity eigenvalues for the time-reversal-invariant momenta (TRIM) points of the bulk Brillouin Zone (BZ) \cite{fu2007topological}, gives a Z$_2$ invariant of -1 for both structures. (We counted 11 electrons for each Au and 4 for Pb.) In both cases the parity at $\Gamma$ is opposite from the parity at all other TRIM leading to Z$_2$ = -1. In the cubic high temperature phase, there is an allowed crossing along the $\Gamma$ - X line that is protected by $C_4$ rotation symmetry (The point group along this line is $C_{4v}$, allowing for two different irreducible representations, $\Gamma_6$ and $\Gamma_7$). In contrast, the potential cones along $\Gamma$ - K are gapped out because the symmetry along this line is $C_{2v}$, which allows only one irreducible representation in the presence of SOC. Thus, around the Fermi level,there are massless Dirac electrons present in high temperature Au$_2$Pb (Figures 4(a) and (4b))- there are also electron and hole pockets at different places within the Brillouin zone. Thus Au$_2$Pb, along with PtBi$_2$, which has a 3D Dirac cone 300 meV below the Fermi energy \cite{gibson20143d}, is one of the very few cubic materials predicted to have a 3D  Dirac cone. Upon cooling into the low temperature structure, the Dirac electrons in Au$_2$Pb become massive because the symmetry reduction allows the bands along $\Gamma$ - X to mix. Nonetheless the low temperature phase is topologically non - trivial. The Z$_2$ = -1 invariant indicates that it must have topological surface states, which should project to the $\bar\Gamma$ point of the surface Brilouin zone, no matter what surface is studied. Note that low temperature Au$_2$Pb is not strictly speaking a TI since the Fermi surface has electron and hole pockets. Due to the continuous gap present in the electronic structure, however, it can be viewed as an insulator with bent bands. The same is observed in elemental Sb which is a topological non trivial metal \cite{teo2008surface}.  With single crystal diffraction we identified the (110), (100) and (111) surfaces of the cubic structure to be exposed in the as-grown material. Spectroscopy studies on the (100) surface, which contains the projection of the $\Gamma$-X line should therefore be feasible. In addition, magneto transport measurement could be undertaken with an applied field along the (100) plane, which theoretically breaks the Dirac point into two Weyl nodes.


We have presented the first example of a real material where a naturally occurring symmetry change gaps out a 3D Dirac cone. In addition to therefore undergoing a massless to massive quasiparticle transition upon cooling, the low temperature phase of Au$_2$Pb is topologically non trivial and becomes superconducting. A few other materials with topological band structures have been shown to be superconducting. These are Cu$_x$Bi$_2$Se$_3$ \cite{hor2010superconductivity}, Sn$_{1-x}$In$_x$Te \cite{balakrishnan2013superconducting}, TlBiTe$_2$ \cite{PhysRevLett.105.266401} and the half Heusler alloys YPtBi \cite{butch2011superconductivity}, ErPdBi \cite{pan2013superconductivity}, LuPdBi \cite{xu2014weak} and LuPtBi \cite{tafti2013superconductivity}. Each of these has disadvantages. In Cu$_x$Bi$_2$Se$_3$ and Sn$_{1-x}$In$_x$Te the superconductivity emerges from a parent TI material but requires doping to induce metallicity. The resulting superconductivity is inhomogenous, true bulk superconductivity is not obtained. The half Heusler compounds do not have a continuous gap, and, in order to display surface states, strain is needed. Au$_2$Pb in contrast has a continuous gap throughout the Brillouin zone in the superconducting regime and is both superconducting and topological at its native composition. In Au$_2$Pb, there is a natural interface between a spin polarized topological surface state and a superconductor in a bulk material.

Topological superconductivity is associated with quasiparticle excitations that are Majorana Fermions \cite{leijnse2012introduction}. It has been proposed that Majorana Fermions can be observed in p-wave superconductors \cite{read2000paired}, where the triplet pairing allows for equal spin projections, or in s-wave superconductors if the electrons in the normal state obey a Dirac like equation \cite{Jackiw1981681}. The second possibility, relevant to the case of Au$_2$Pb, is predicted to occur at the surface of topological insulators (TIs) by using the proximity effect to induce superconductivity in those materials \cite{fu2008superconducting}. The strong spin splitting of the TI allows for p-wave Cooper pairing on the surface in spite of the s-wave superconductivity in the bulk. The Fermi energy must be in the band inverted regime of the electronic structure. In Au$_2$Pb there is a natural interface between a spin polarized topological surface state and a superconductor in a bulk material. Because the T$_c$ of Au$_2$Pb is high enough for STM studies, this compound is a uniquely promising candidate for studying the interaction between superconductivity and topological surface states. 

In addition, the phase transition observed in Au$_2$Pb allows for studies of the breakdown of the 3D Dirac cone on symmetry breaking. It has been proposed that the known 3D Dirac semimetals Cd$_2$As$_3$ and Na$_3$Bi should transition into a TI state if they are distorted \cite{wang2012dirac,wang2013three}, but this will be difficult to achieve experimentally as such distortions are not naturally occurring in these materials. Au$_2$Pb, however, distorts naturally upon cooling and is therefore eligible for those studies. Finally, we note that the electronic structure of cubic Au$_2$Pb does not only have the 3D Dirac cone at the Fermi level but also other bands crossing on the $\Gamma$-K line. The low temperature electronic structure is topologically non trivial but has a Fermi surface composed of electron and hole pockets. These pockets are what allows the compound to become superconducting. This is in contrast to a true TI which cannot become superconducting without doping. As such, our work on Au$_2$Pb indicates that 3D Dirac materials that have other pockets that are well separated away in the Brillouin zone should be an interesting point of future research, as more instabilities and symmetry breaking transitions are possible.

\section{Methods}
Single crystals of Au$_2$Pb were grown out of Pb flux. Au (99.999\% purity) and Pb (99.9999\% purity) were mixed in the molar ratio 40:60 and heated in an evacuated quartz tube to 600$^\circ$C for 24 h then slowly cooled (3$^\circ$/min) to 300$^\circ$C. At this temperature the flux was separated from the crystals. Excess Pb was removed from the crystals by etching them in an aqueous acetic acid / hydrogen peroxide solution for one minute. Single crystal x-ray diffraction data were collected on a Bruker Kappa APEX II diffractometer using graphite-monochromated Mo K$\alpha$ radiation ($\lambda$=0.71073 \AA) at 100~K. Unit cell determination and refinement, and data integration were performed with Bruker APEX2 software. The crystal structure was determined using SHELXL-97 \cite{Sheldrick:sc5010} implemented through WinGX \cite{farrugia1999wingx}. Synchrotron X-ray diffraction studies were performed on Au$_2$Pb at the Advanced Photon Source at Argonne National Laboratory on beamline 11BM. The patterns were collected from 7 K to 154 K in 2.5 K increments, with a wavelength of 0.413841 \AA~ and a 2$\theta$ range of 0.5 degrees to 26 degrees. Structural solution was performed by utilizing distortion mode analysis by utilizing a combination of ISODISTORT and TOPAS \cite{campbell2007rietveld,Campbell:wf5017}. The powder data were refined with the Rietveld method, using the FULLPROF program \cite{Juan199355}. Resistivity and heat capacity measurements were performed with a Physical Property Measurement System (PPMS) from Quantum Design, equipped with a $^3$He cryostat.

Electronic structure calculations were performed in the framework of density functional theory (DFT) using the \textsc{wien2k} \cite{blaha2001} code with a full-potential linearized augmented plane-wave and local orbitals [FP-LAPW + lo] basis \cite{singh2006} together with the Perdew-Becke-Ernzerhof (PBE) parameterization \cite{perdew_generalized_1996} of the Generalized Gradient Approximation (GGA) as the exchange-correlation functional. The plane wave cut-off parameter R$_{MT}$K$_{MAX}$ was set to 8 and the irreducible Brillouin zone was sampled by 560 k-points (cubic) and 216 k-points (orthorhombic). Experimental lattice parameters from the Rietveld refinements were used in calculations. Spin orbit coupling (SOC) was included

\bigskip 
\begin{acknowledgments}
This research was supported by the ARO MURI on topological insulators, grant W911NF-12-1-0961. This research used resources of the Advanced Photon Source, a U.S. Department of Energy (DOE) Office of Science User Facility operated for the DOE Office of Science by Argonne National Laboratory under Contract No. DE-AC02-06CH11357. We acknowledge support from a MRSEC grant from the US National Science Foundation DMR 0819860. This work is also supported by the Laboratory Directed Research and Development Program and the Molecular Foundry of Lawrence Berkeley National Laboratory under DOE Contract No. DE-AC02-05CH11231. We thank NERSC for computational resources.
\end{acknowledgments}

\section{Author Contribution}
LMS is the lead researcher. She and LSX grew crystals, calculated electronic structures and measured low temperature resistivities and heat capacity with NH. RC and IK performed detailed theoretical analysis including determining the Z$_2$ invariants. SHL performed low temperature diffraction experiments and solved low temperature structure. MH, CAB and TL measured the detailed resistivity behavior. MNA performed single crystal diffraction studies. QDG helped analyzing results and putting them into context. JBN, NPO, AV  and RJC supervised the research. All authors contributed to writing the manuscript.
%

\clearpage

\begin{figure}[htb]
  \centering
  \includegraphics[width=13cm]{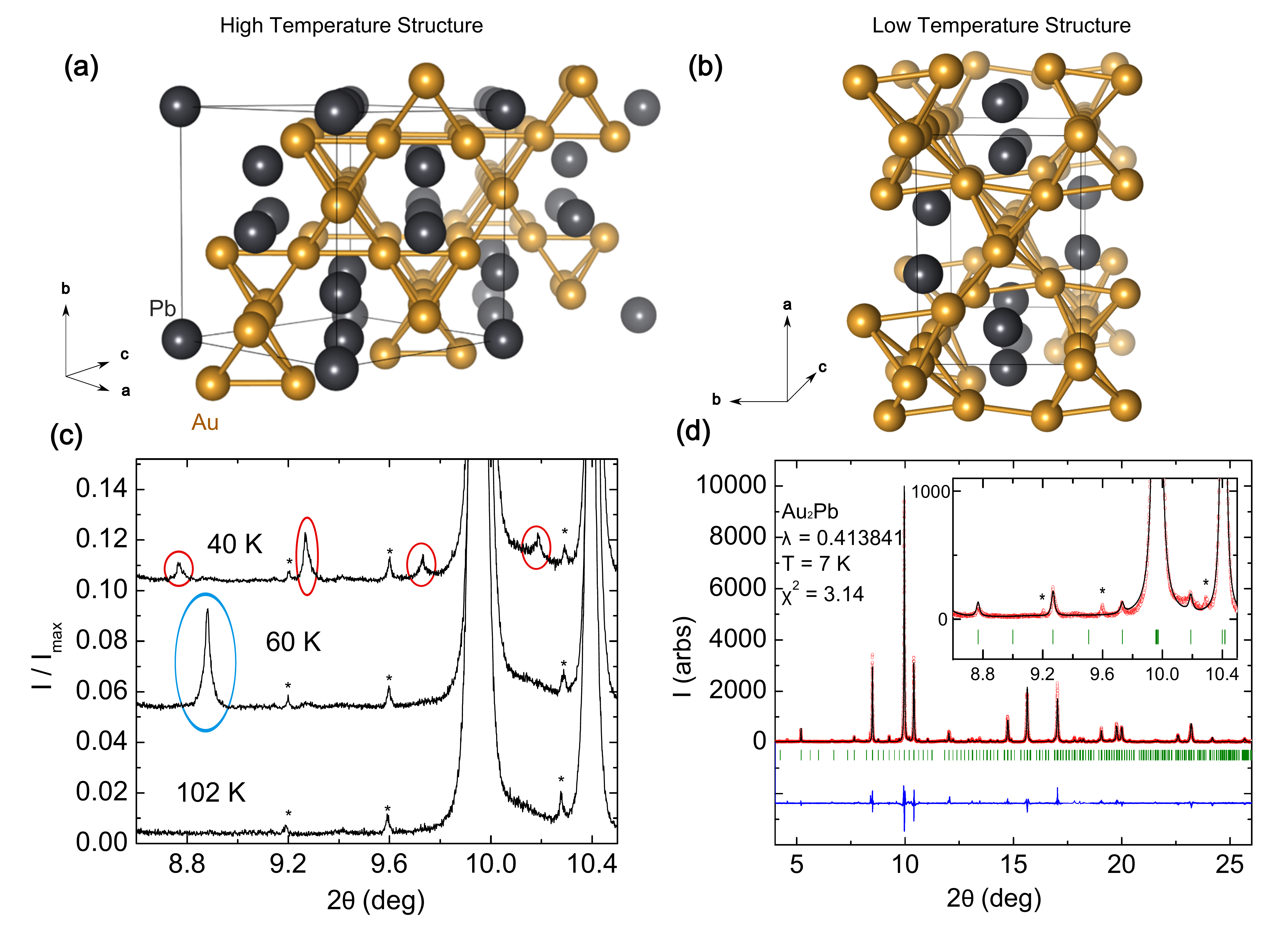}
  \caption{Structural information on Au$_2$Pb. (a) Crystal structure of the high temperature cubic Laves phase Au$_2$Pb.(b) Crystal structure of the low temperature orthorhombic, distorted phase of Au$_2$Pb. (c) Synchrotron diffraction patterns at different temperatures showing the presence of three different phases in Au$_2$Pb. Newly appearing peaks are circled, peaks associated with the orthorhombic phase are circled in red. Blue circled peaks belong to the intermediate phase. Impurity phases are marked with asterisks; those peaks remain present across the whole temperature regime measured. Patterns are displaced for clarity. (d) Rietveld refinement of orthorhombic Au$_2$Pb at 7 K, the insert shows the region shown in (c); all the newly appearing peaks are captured by the structural model.}
\label{struc}
\end{figure}

\clearpage

\begin{figure}[htb]
  \centering
  \includegraphics[width=18cm]{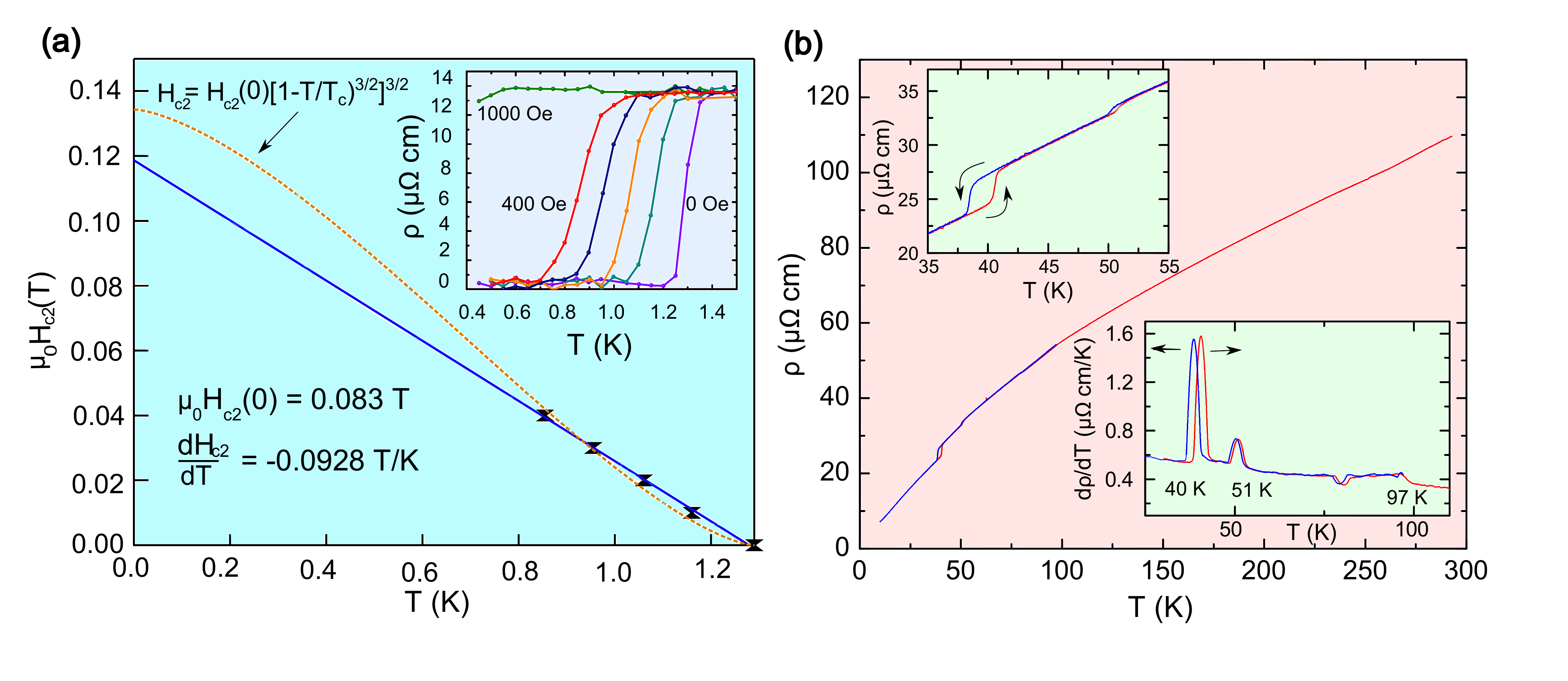}
  \caption{Characterization of the superconductivity and normal state resistivity in Au$_2$Pb. (a) Critical field determination of Au$_2$Pb, the blue line shows the linear fit to determine the slope $dH_{c2}/dT$ = - 0.0928 T/K. The orange line is a fit to the Ginzburg Landau Theory with the formula $H_{c2} = H_{c2}(0) [1-(T/T_c)^{3/2}]^{3/2}$. The inset shows the resistivity close to T$_c$ for different fields. (b) $\rho$(T) between 300 and 25 K. Three discontinuities can be seen, at 97 K, 51 K and 40 K. The lower inset shows the derivative which shows the discontinuities more clearly. Blue lines represent data measured on cooling while red lines represent data measured on heating.}
\label{RT}
\end{figure}
\clearpage
\begin{figure}[htb]
  \centering
  \includegraphics[width=13cm]{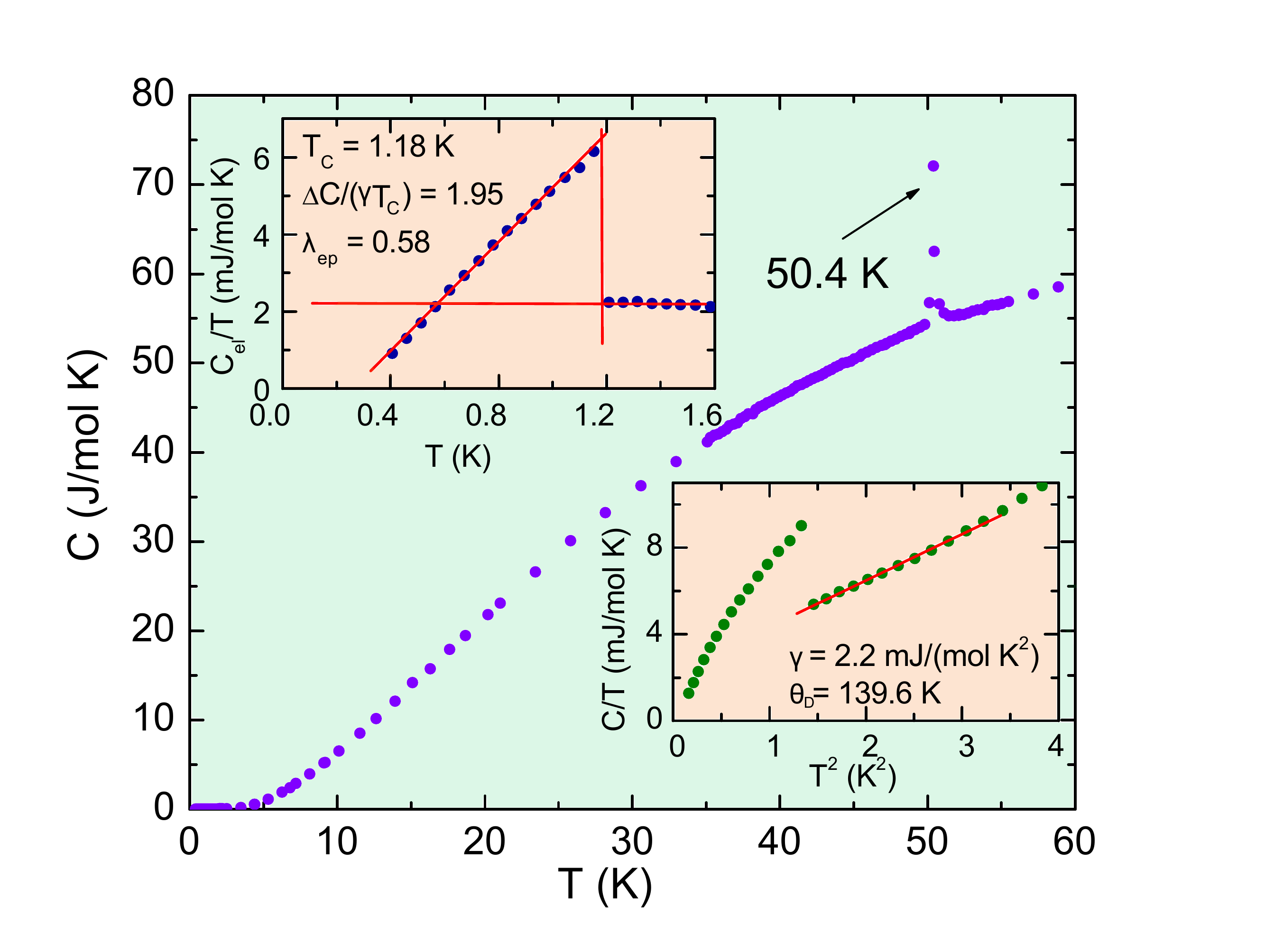}
  \caption{Specific heat of Au$_2$Pb, confirming the bulk superconductivity and a phase transition at 50.4 K. The upper inset shows the electronic specific heat vs. temperature of Au$_2$Pb. The sharp jump indicates bulk superconductivity. The value of $\Delta C/(\gamma \cdot T_c) = 1.95$ is higher than the expected BCS value. The lower inset shows the linear fit above T$_c$ that was used to determine the Sommerfeld coefficient $\gamma$ and the coefficient for the phononic contribution to the heat capacity $\beta$.}
\label{heatCap}
\end{figure}

\clearpage

\begin{figure}[htb]
  \centering
  \includegraphics[width=13cm]{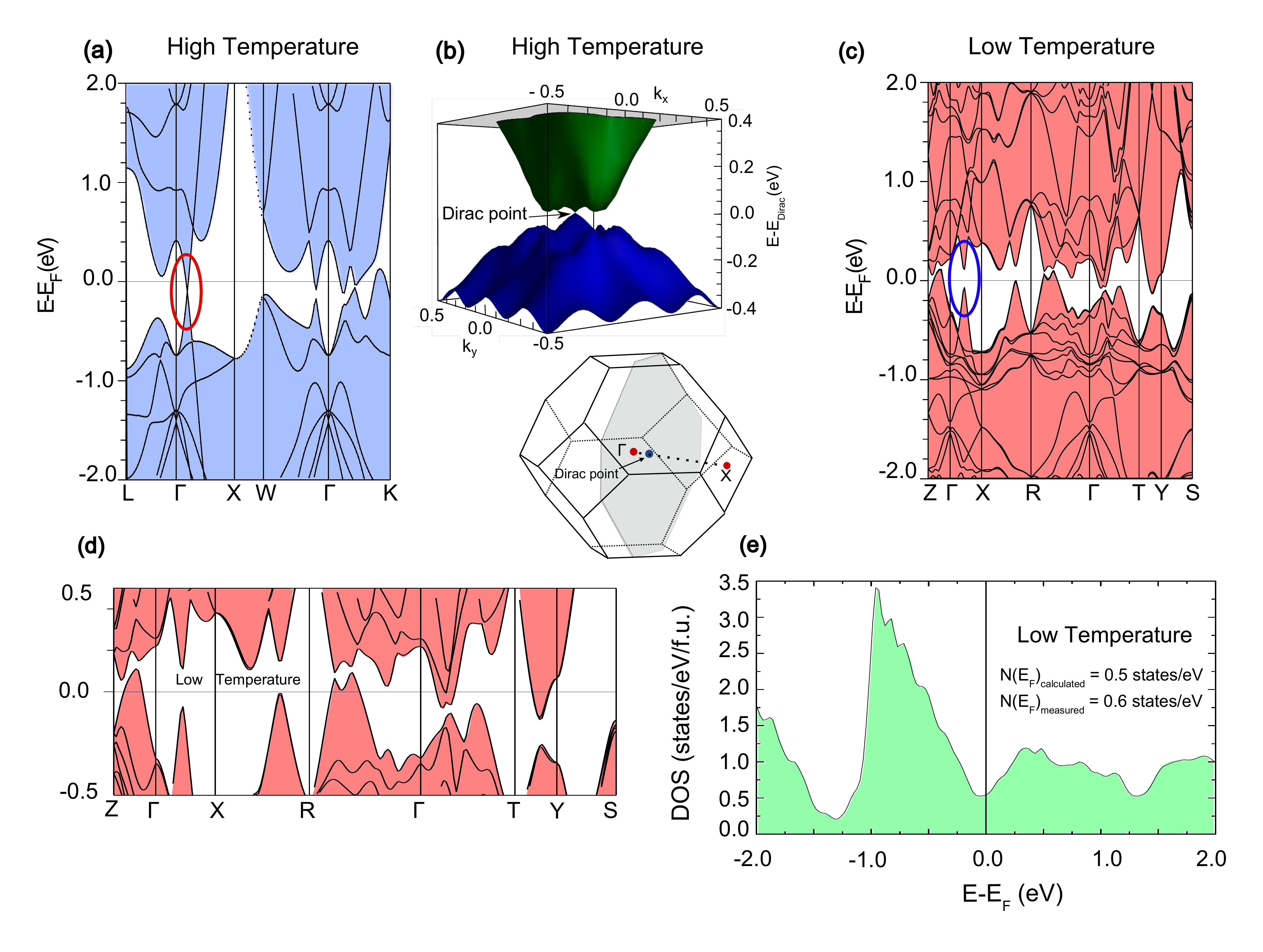}
  \caption{Electronic structures of Au$_2$Pb. Band structure of high temperature cubic (a) and low temperature orthorhombic (c) Au$_2$Pb .Panel (b) shows a view of the electronic structure of the high temperature phase along the (100) direction. (d) emphasizes the continuous gap in the electronic structure of orthorhombic Au$_2$Pb. An energy dependent density of states (DOS) plot is shown for orthorhombic Au$_2$Pb in (e). The cubic electronic structure has a 3 D Dirac cone along $\Gamma$-X (circled), which gets gapped out when the compound becomes distorted at low temperatures.  Despite the metallic pockets the Dirac point should be visible in APRES measured on the (100) surface. Orthorhombic Au$_2$Pb has a continuous gap and can be viewed as a semiconductor with bent bands. The calculated DOS at E$_F$ matches well with the measured value.}
\label{bands}
\end{figure}
\clearpage

\begin{table}
\caption{(S1) Structural information for low temperature (T = 6.76 K) orthorhombic Au$_2$Pb. Space group \textit{Pbnc}, a = 7.9040(7) \AA ~ b = 5.5786(10) \AA ~ c =11.1905(2) \AA. }
\begin{center}
\begin{tabular}{ccccc}
\hline
\hline 
\textbf{Atom} & \textbf{Wyckoff} & \textbf{x} & \textbf{y} & \textbf{z}\\[1ex]
Pb & 8d  & 0.3696(2) & -0.0062(2) & 0.1274(2)\\
Au 1 & 4c & 0.0000 & 0.0120(4) & 0.25000\\
Au 2 & 8d & 0.2741(1) & 0.2148(2) & 0.3742(2)   \\
Au 3 & 4a  & 0.00000 & 0.00000 & 0.00000 \\
\hline
\hline
\end{tabular}
\end{center}
\label{structure}
\end{table}

\begin{table}
\caption{(S2) Superconducting properties of Au$_2$Pb.}
\begin{center}
\begin{tabular}{cc}
\hline
\hline 
\textbf{Parameter} & \textbf{Au$_2$Pb}  \\[1ex]
T$_c$ & 1.18 K   \\
$dH_{c2}dT$ &  - 0.0928 T/K \\
$\mu_0H_{c2}$ & 83 mT\\
$\xi(0)$ & 629 \AA \\
$\gamma$ & 2.2 mJ/mol$\cdot$ K$^2$  \\ 
$\frac{\Delta C}{\gamma T_c}$ & 1.95  \\
$\Theta_D$ & 139.6 K    \\  
$\lambda_{ep}$ & 0.58   \\
N(E$_{F}$)$_{measured}$ & 0.6 states/eV/f.u.   \\ 
N(E$_{F}$)$_{calculated}$ & 0.5 states/eV/f.u.  \\
\hline
\hline
\end{tabular}
\end{center}
\label{prop}
\end{table}

\end{document}